\documentclass[11pt,a4paper]{article}
\usepackage{jcappub}
\usepackage{graphicx}

\title{Cosmological evidence for leptonic asymmetry after {\sc Planck}}

\author{A. Caramete and L. A. Popa \footnote{Corresponding author: L. A. Popa $<$lpopa@spacescience.ro$>$} }

\affiliation{Institute of Space Science,\\
Bucharest-Magurele, Ro-077125 Romania}

\emailAdd{acaramete@spacescience.ro, lpopa@spacescience.ro}

\abstract{ Recently, the {\sc Planck} satellite found a larger and most precise value of
the matter energy density, that impacts on the present values of
other cosmological parameters such as the Hubble constant $H_0$, the present cluster abundances $S_8$, and the age of the Universe $t_U$.
The existing tension between {\sc Planck} determination of these parameters in the frame
of the base $\Lambda$CDM model and their determination from other measurements generated lively discussions,
one possible interpretation being that some sources of systematic errors in cosmological measurements are not completely understood. \\
An alternative interpretation is related to the fact that the CMB observations, 
that  probe the high redshift Universe are interpreted in terms of cosmological parameters at present time  by extrapolation within the base $\Lambda$CDM model that can be inadequate or incomplete. \\
In this paper we quantify this tension by exploring several extensions of the base $\Lambda$CDM model that include the leptonic asymmetry. We set bounds on the radiation content of the Universe and neutrino properties by using the latest cosmological measurements, imposing also self-consistent BBN constraints on the primordial helium abundance. 
For all asymmetric cosmological models we find the preference of cosmological data for smaller values of active and sterile neutrino masses. This increases the tension between cosmological and short baseline neutrino oscillation data
that favors a sterile neutrino with the mass of around 1~eV.
For the case of degenerate massive neutrinos, we find that the discrepancies with the local determinations of $H_0$, and $t_U$ are alleviated at $\sim 1.3\sigma$ level while $S_8$ is in agreement with its determination from CFHTLenS survey data at $\sim 1 \sigma$ and with the prediction of cluster mass-observation relation at $\sim 0.5\sigma$. \\
We also find $2 \sigma$ statistical preference of the cosmological data 
for the leptonic asymmetric models involving three massive neutrino species and neutrino direct mass hierarchy.
We conclude that the current cosmological data favor
the leptonic asymmetric extension of the base $\Lambda$CDM model and normal neutrino mass hierarchy over the models  with additional sterile neutrino species and/or inverted neutrino mass hierarchy.
}


\begin{document}
\maketitle
\begin{flushright}
\end{flushright}

\section{Introduction}

The cosmological observations have established the minimal flat $\Lambda$CDM model as standard model for cosmology. With six basic parameters, $\Omega_bh^2, \Omega_ch^2, \tau, \theta_A, A_S, n_S $
(where $\Omega_b h^2$ is the baryon energy density, $\Omega_ch^2$ the cold dark matter energy density, $\tau$ is the Thomson optical depth to reionization, $\theta_A$ is the angular acoustic scale at recombination and $A_S$ and $n_s$ are the amplitude and the spectral index of the initial scalar perturbations respectively), this base model  can explain
the acoustic Doppler peaks structure of the Cosmic Microwave Background (CMB) angular power spectra, the large scale structure (LSS) formation via gravitational instability, the abundance of clusters at small redshifts, the spatial distribution and the number density of galaxies, the expansion history of the Universe and the cosmic acceleration.

Recently, the {\sc Planck} satellite \cite{Planck1} established an excellent agreement
between the power spectra of the CMB temperature anisotropies at high multipoles and of 
the lensing potential with the predictions of the base $\Lambda$CDM cosmological model \cite{Planck2}. {\sc Planck} 
found quite large changes in some parameters of the base $\Lambda$CDM model 
when compared with those from other astrophysical measurements.
In particular, {\sc Planck} found a larger and most precise value of
the matter energy density, $\Omega_m$, that impacts on the present values of
other cosmological parameters such as the Hubble constant, $H_0$, and the cluster abundances.\\
The lower value of $H_0$ found by {\sc Planck} in the frame of the
base $\Lambda$CDM model is consistent (within 1$\sigma$) with the value of $H_0$ obtained by the WMAP experiment \cite{Hinshaw2012} but in tension (at about 2.5$\sigma$)
with $H_0$ local measures \cite{Reiss2011,Freedman2012}.
Also, {\sc Planck} found an increased value of the local cluster abundance which is 
in significant tension (at about 3$\sigma$) with similar values reported by other analysis
\cite{Benson2013,Reichardt2013,Hasselfield2013}, 
including the analysis of {\sc Planck} cluster counts \cite{Ade1_2013}. \\ 
Since clusters provide estimates of the cluster mass normalization condition, this uncertainty is 
dominated by the impact of $\Omega_m$ on the  growth function, but also depends on other parameters 
such as neutrino mass. Recent works \cite{Rozo2013,Wyman} illustrate how 
a more accurate cluster mass-observable relation in determining the cluster mass normalization  
impacts on the neutrino mass determination, contributing to alleviate 
this tension significantly.

Following the {\sc Planck} team suggestion \cite{Planck2}, 
one possible interpretation of the existing tension is that 
some sources of systematic error in cosmological measurements are not completely understood.
The tension found between  different datasets has been also discussed in Ref. \cite{Hou2012}. 

An alternative interpretation of this tension is related to the fact that the CMB observations, that  probe the physics of early Universe (up to redshift of $z \sim$ 1000 or $\sim$ 380,000 years after the Big Bang), are interpreted in terms of cosmological parameters at present time ($z$ = 0) by extrapolation within the base $\Lambda$CDM model that can be inadequate or incomplete. 
Extensions of the baseline $\Lambda$CDM model has been recently explored by including several extra parameters that can alleviate the tension. The analysis presented Ref. \cite{Hou2012} shows that, for all models that the authors considered, no single parameter extension to the baseline $\Lambda$CDM model helps to alleviate the tension. \\
An interesting finding \cite{Verde2013}
is that when using the neutrino mass as additional parameter to extend the base $\Lambda$CDM model,  a total neutrino mass above $0.15$ eV makes the tension highly significant, showing that
the degenerate neutrino hierarchy is highly disfavored by the data.\\
The extension of the base $\Lambda$CDM model by the
inclusion of an extra radiation energy density (besides photons)
due to relativistic species in terms of neutrino temperature, usually parametrized by the effective number of relativistic degrees of freedom $N_{eff}$ \cite{Giunti2007,Lesgourgues2012,Lesgourgues2013},
has been also extensively discussed \cite{Wyman,Verde2013,Jaques2013,Bari2013,Boem2013,Kelso2013,Valentino,Said,Weinberg,Peiris,Hamann,Battye,Ri2013a,Ri2013b}.
Since the value of $N_{eff}$ in the Sandard Model (SM) is $N_{eff}$=3.046 \cite{Mangano},  the detection of any positive deviation from this value would be a signal
that the radiation content of the Universe was due not only to photons and neutrinos, but also to some additional relativistic relics. The determination of $N_{eff}$ from cosmological data is closely related to the determination of $H_0$.
Since $N_{eff}$ and $H_0$ are positively correlated, the tension between the {\sc Planck} data and local measures of $H_0$  can be relieved in the base $\Lambda$CDM model for
$N_{eff}$ values  around 3.6-3.8 \cite{Verde2013,Wyman,Gari2013}. However, the constraints on $N_{eff}$ obtained by the {\sc Planck} team show no strong preference of data for the existence of the extra relativistic degrees of freedom \cite{Planck2,Peiris}.\\
The extension of the base $\Lambda$CDM model by including a sterile neutrino with mass
in the eV range, as suggested by the short baseline and reactor neutrino oscillation anomalies \cite{Aguilar2012,Giunti2013}, has also been considered \cite{Wyman,Gari2013,Feeney,Archid2013,Krist2013}. 
The addition of a sterile neutrino to three massive active neutrinos simultaneously change the acoustic scale and suppress the growth of structures, bringing the measures of $H_0$ and cluster abundances closer to their determinations from cosmological data. Other  extensions of the base $\Lambda$CDM model include panthom values of dark energy equation of state ($w \sim -1.2$) or a small positive curvature ($7 \times 10^{3}<\Omega_K <1.5 \times 10^{-2}$) \cite{Verde2013}.

The goal this paper we quantify the existing tension between the {\sc Planck} data and other astrophysical measurements by exploring several extensions of the base $\Lambda$CDM model that include the leptonic asymmetry.
The existence of a large leptonic asymmetry is restricted to be in the form of neutrinos from the requirement of universal electric neutrality.
Although the Standard Model predicts the leptonic asymmetry of the same order as the baryonic asymmetry, $L \simeq B\sim 10^{-10}$, there are particle physics scenarios in which much larger leptonic asymmetry can be produced \cite{Smith2006,Cirelli2006,Kirilova2013}. \\
The leptonic asymmetry  is most conveniently measured by the neutrino degeneracy parameter
\cite{Dolgov,Serpico,Simha,Popa2008} defined as
$\xi_{\nu}=\mu_{\nu} /T_{\nu,0}$, where $\mu_{\nu}$ is the neutrino chemical potential and
$T_{\nu,0}$ is the present temperature of the neutrino background [$T_{\nu,0}/T_{CMB}=(4/11)^{1/3}$].
As the measured neutrino mixing parameters imply that the
active neutrinos reach the chemical equilibrium before Big Bang Nucleosynthesis
(BBN) \cite{Wong,Abazajian}, in this paper we consider tree massive neutrino flavors $\nu_{\alpha}$ ($\alpha= e, \mu, \tau$) with degenerated chemical potential, $\xi_{e}=\xi_{\mu}=\xi_{\tau}$. \\
The most important impact of the leptonic asymmetry is the increase of the radiation energy density that, for three neutrino species with degenerated chemical potential $\xi_{\nu}$, can be parametrized by:
\begin{equation}
\label{delta_neff_zeta}
\Delta N_{eff} (\xi_{\nu})= 3 \left[\frac{30}{7} \left(\frac{\xi_{\nu}}{\pi}\right)^2
+\frac{15}{7} \left(\frac{\xi_{\nu}}{\pi}\right)^4 \right] \,.
\end{equation}
The radiation extra energy density can then be splitted in two uncorrelated  contributions:
\begin{equation}
\label{delta_neff}
\Delta N_{eff} = \Delta N_{eff}(\xi)+ \Delta N^{oth}_{eff} \,,
\end{equation}
first due to the net leptonic asymmetry of the neutrino background and second due to the extra contributions from other unknown processes.
The radiation extra energy density delays the time of matter-radiation equality, boosting the acoustic Doppler peaks of the CMB power spectrum. For the same reasons the acoustic peaks are shifted to higher multipoles.\\
Also, the temperature anisotropy of the neutrino background (the anisotropic stress) that acts as an additional source term for the gravitational potential, changes the CMB anisotropy power spectrum at the level of ∼ 20$\%$ \cite{Hu95,Trotta2005}.
The delay of the epoch of matter-radiation equality shifts the
matter density fluctuations power spectrum turnover position toward larger angular scales, suppressing the power at small scales. In particular, the non-zero neutrino chemical potential leads to changes in neutrino free-streaming length and neutrino Jeans mass due to the increase of the neutrino velocity dispersion \cite{Pastor1999,Latanzzi,Ichiki}. \\
The leptonic asymmetry also shifts the beta equilibrium between protons and neutrons at the BBN epoch, leading to indirect effects on the CMB anisotropy through the primordial helium abundance, $Y_P$, that decreases monotonically with increasing $\xi_{e}$.\\
Details of the effects of the neutrino mass and leptonic asymmetry on BBN and CMB can be found in Refs. \cite{Kinney,Ichiki,Popa2008,Ichikawa}.

This paper is organized as follows. In Sec. 2 we describe the methods used in our analysis and the datasets and combinations of datasets we consider.
We present our results in Sec. 3 where we examine the consistency and cosmological implications
of our results obtained in the leptonic asymmetric cosmological models for degenerate and hierarchical neutrino masses and in Sec. 4 we draw our conclusions.

\section{Model, methods and datasets}

The density perturbations in leptonic asymmetric cosmological models have been
discussed several times in literature \cite{Pastor1999,Latanzzi,Ichiki,Kinney,Popa2008,Ichikawa}. We applied them to modify the Boltzmann Code for Anisotropies in the Microwave Background, CAMB \footnote{\url{http://camb.info/}} \cite{camb}, to compute the
CMB temperature and polarization anisotropies power spectra and matter density
fluctuations power spectra for the case of three massive neutrinos/antineutrinos with
the total mass $\Sigma m_{\nu}$ and degeneracy parameter $\xi_{\nu}$ . As neutrinos reach their approximate
chemical potential equilibrium before BBN epoch \cite{Dolgov,Wong,Abazajian}, we consider in our
computation that all three neutrino/antineutrino flavors have the same degeneracy
parameter $\xi_{\nu}$.
We modify the expressions for neutrino/antineutrino density and pressure in
the relativistic and non-relativistic limits and follow the
standard procedure to compute the perturbed quantities by expanding the phase space
distribution function of neutrinos and antineutrinos into homogeneous and perturbed
inhomogeneous components. Since the gravitational source term in the
Boltzmann equation is proportional to the logarithmic derivative of the neutrino/antineutrino distribution functions with respect to comoving momentum, we also modify this term to account for $\xi_{\nu} \ne 0$ \cite{Pastor1999,Popa2008,Ichiki}. 

For our cosmological analysis we use a modified version of the latest publicly available package CosmoMC \footnote{\url{http://cosmologist.info/cosmomc/}}
\cite{Lewis02} to sample
from the space of possible cosmological parameters and generate estimates of the posterior mean of each parameter of interest and the confidence interval.\\
We use the following datasets and likelihood codes:
\begin{itemize}
\item The first public release of {\sc Planck} Collaboration temperature data, combined with WMAP-$9$ year polarization information at low $\ell$, and the corresponding likelihood codes \cite{Planck1,PlanckXV}:
\texttt{Commander}, that computes the low-$l$ {\sc Planck} likelihood,
\texttt{CamSpec}, that computes the {\sc Planck}  likelihood for the multipoles with $50\leq l\leq 2500$,
\texttt{LowLike}, that computes the likelihoods from the $2\leq l\leq 32$ temperature and polarization data
\footnote {Since there are no {\sc Planck} polarization data in this first cosmological data release, WMAP polarization data \cite{WMAP9} are used to constrain the optical depth to reionization, $\tau$.}
and \texttt{Lensing}, that computes the likelihoods from {\sc Planck} lensing power spectrum data, for multipoles between 40 and 400 \cite{PlanckXVII}.
\item High-$l$ data from ground-based telescopes Atacama Cosmology Telescope(ACT) \cite{ACT13,Dunkley13} and the South Pole Telescope (SPT) \cite{Keisler11,Reichardt2013}. These experiments have mapped the foregrounds with higher resolution and lower noise than {\sc Planck} and can complement the {\sc Planck} data to better constrain the foreground-model parameters \cite{Peiris}.
\item Geometrical constraints from baryon acoustic oscillation (BAO). The BAO in the distribution of galaxies are extracted from the most recent redshift survey data: the Sloan Digital Sky Survey (SDSS) Data Release 7 (DR7) \cite{BAODR7} at $z_{eff} = 0.2$ and $z_{eff} = 0.35$, the reanalyzed SDSS DR7 galaxy catalog data at $z_{eff} = 0.35$ \cite{DR7, Padman12}, the SDSS Baryon Oscillation Spectroscopic Survey(BOSS) Data Release 9 (DR9) at $z_{eff} = 0.57$ \cite{DR9,BAODR9} and the 6dF Galaxy Survey (6dFGS) \cite{6dF} measurements at $z_{eff}=0.1$ \cite{6dF, BAO6dF}.
\item BBN prediction of the helium abundance, $Y_P$, for different values of $\Omega_b h^2$, $\Delta N_{eff}$ and $\xi_{\nu}$ used in the analysis, as given by the BBN PArthENoPE code \cite{Pisanti07, Kneller04}.
\end{itemize}
The full dataset used in our analysis is summarized in Table~\ref{tab:date}.
The CosmoMC code uses also other 31 extra parameters, to account for the foreground and nuisance parameters of {\sc Planck} and ACT/SPT data that are described in detail inside the code current distribution.
\begin{table}[tb]
\caption{Summary of the datasets, their abbreviation and relevant references.}

\vspace{0.2cm}
\centering
\begin{tabular}{lcc}
\hline
Data & Abbreviation & Reference\\
\hline {\sc Planck} temperature & {\sc Planck} & \cite{PlanckXV}\\
WMAP low $\ell$ polarization  & WP &\cite{WMAP9}\\
{\sc Planck} reconstructed lensing potential & lensing & \cite{PlanckXVII}\\
ACT and SPT& highL& \cite{ACT13, Dunkley13, Keisler11,Reichardt2013}\\
Baryon Acoustic Oscillations compilation& BAO& \cite{DR7, BAODR7, Padman12, DR9, BAODR9, 6dF, BAO6dF}\\
BBN constraints on $Y_P$ &BBN & \cite{Pisanti07, Kneller04}\\
Neutrino mass squared differences& $\nu$MSD & \cite{Maltoni}\\
\hline
\end{tabular}
\label{tab:date}
\end{table}

\section{Analysis}

We consider the following extensions of the base $\Lambda$CDM cosmological model:\\\\
{\bf $m\Lambda$CDM+$\Sigma m_{\nu}$}:
The minimal extention the base $\Lambda$CDM model
by the addition of three species of massive neutrinos with total mass $\Sigma m_{\nu}$
(within which $\Lambda$CDM is nested at $N_{eff}$=3.046 and $Y_P$=0.24).\\
{\bf $m\Lambda$CDM+$\Delta N^{oth}_{eff}$+$m^{eff}_{s}$+$Y_P$}:
The addition to $m\Lambda$CDM model of the extra neutrino species $\Delta N^{oth}_{eff}$ from unknown processes,
one sterile neutrino with the effective mass $m^{eff}_s$ and the BBN prediction of the primordial helium abundance $Y_P$.
The effective sterile neutrino mass $m^{eff}_s$ is related to the
true mass $m_s$ in one of the two ways.
If sterile neutrino is distributed with a temperature $T_s$ which is
different from the temperature $T_{\nu}$ of active neutrinos,
$m^{eff}_s= (T_s/T_{\nu})^3m_s= (\Delta N_{eff})^{3/4}m_s$.
Alternatively, if sterile neutrino is distributed with the same temperature as active neutrinos suppressed by a constant
factor $\chi_s \leq 1$, as in the case of Dodelson-Widrow model \cite{DW},
$m^{eff}_s = \chi_s m_s = \Delta N_{eff}\, m_s$.
Since the two cases are in fact equivalent for the cosmological analysis,
we consider in this work the termally distributed sterile neutrino case. \\
{\bf $m\Lambda$CDM+$\Delta N_{eff}(\xi)$+$\xi_{\nu}$+$Y_P$}:
The addition to $m\Lambda$CDM of the neutrino chemical potential $\xi_{\nu}$,
the extra neutrino species $\Delta N_{eff}(\xi)$
as given in Eq. (\ref{delta_neff_zeta}) and the BBN prediction of the primordial helium abundance $Y_P$.\\
{\bf $m\Lambda$CDM+$\Delta N_{eff}$+$\xi_{\nu}$+$Y_P$}:
The addition to $m\Lambda$CDM of the neutrino chemical potential $\xi_{\nu}$,
the extra neutrino species $\Delta N_{eff}$
as defined in Eq. (\ref{delta_neff}) and the BBN prediction of the primordial helium abundance $Y_P$.\\
{\bf $m\Lambda$CDM+$\Delta N_{eff}$+$\xi_{\nu}$+$m^{eff}_s$+$Y_P$}:
The addition to the previous model of one sterile neutrino termally distributed
with the effective mass $m^{eff}_s$. \\
From the analysis of fundamental Monte Carlo Markov Chain (MCMC) parameters we obtain
the posterior probability distributions for the Hubble constant $H_0$, the age of the Universe $t_U$,
and the cluster mass normalization $S_8$. We then compare the cosmological constraints on these parameters with 
their corresponding astrophysical measurements. 

For Hubble parameter, $H_0$, we use two most recent reported local measurements \cite{Reiss2011,Freedman2012}.
The average of these two values with the central value given by variance-weighted mean and error
given by the average of errors \cite{Verde2013} is:
\begin{eqnarray}
\label{H0}
H_0=74.08 \pm 2.25\,\, {\rm km s^{-1} Mpc^{-1}}\,.
\end{eqnarray}
We also use the local measures of the age of the Universe, $t_U$, 
obtained from the ages of the oldest stars, since these objects form very shortly after the Big Bang. 
Accurate dating, based on accurate 
distance determination using direct parallax measurements were obtained for nearby sub-giant stars. 
In particular, the age of the nearby sub-giant HD-140283 determined to be $14.5 \pm 0.8$ Gyr has been accurately measured by using the HST parallaxes and spectroscopic determinations of its chemical abundance \cite{Bond2013}. Additionally,
the ages for some of the most metal poor Milky Way globular clusters have been determined 
to be $14.2 \pm 0.6$ ($\pm 0.8$ systematics) Gyr \cite{Imb2004,Gratton}. 
As the ages determination of nearby sub-giants and globular clusters are
dominated by different and independent systematics,  
the combination of the two above measurements by inverse variance weighting 
lead to the following estimate of the age of Universe \cite{Verde2013}:
\begin{equation}
\label{tU}
t_U=14.61 \pm 0.8 \,\,{\rm Gyr}\,.
\end{equation}
The advantage of using the local values of $H_0$ and $t_U$ is that 
they are measured at $z=0$ and therefore no cosmology-dependent extrapolation is needed. \\
We take the value of $S_8$ obtained from the estimates of shear correlation functions associated with
six redshift bins of the most recent and largest shear dataset provided by the CFHTLenS survey \cite{Heymans}:
\begin{eqnarray}
\label{S8}
S_8=\sigma_8 ( \Omega_m / 0.27 )^{0.46}=0.774 \pm 0.04\,.
\end{eqnarray}
One should note that the above value of $S_8$ is obtained assuming a flat $\Lambda$CDM model. \\
We also employ the cluster mass-observable relation  that gives the  dependence of  $S_8$
on cluster mass calibration in the form \cite{Vik2009}:
\begin{eqnarray}
S_8=S_{8,V09} + 0.024 \frac {\Delta {\rm ln} M } {0.09}\,, 
\hspace{0.6cm}S_{8,V09}=\sigma_8 ( \Omega_m/0.25)^{0.47}=0.813  \,,
\end{eqnarray}
where $\Delta {\rm ln}M$ is the cluster mass calibration offset relative to  $S_{8,V09}$. \\
We consider two mass calibration offset values,  
${\rm P11}=-0.12 \pm 0.02$ adopted by {\sc Planck} team \cite{Ade2011} 
and ${\rm R12}=0.11 \pm 0.04$ corresponding to the cluster mass scale employed in Ref. \cite{Rozo2012b}. 

\subsection{Consistency and cosmological implications}

\subsubsection{Degenerate massive neutrinos}

\begin{figure}
\begin{center}
\includegraphics[height=14cm,width=14cm]{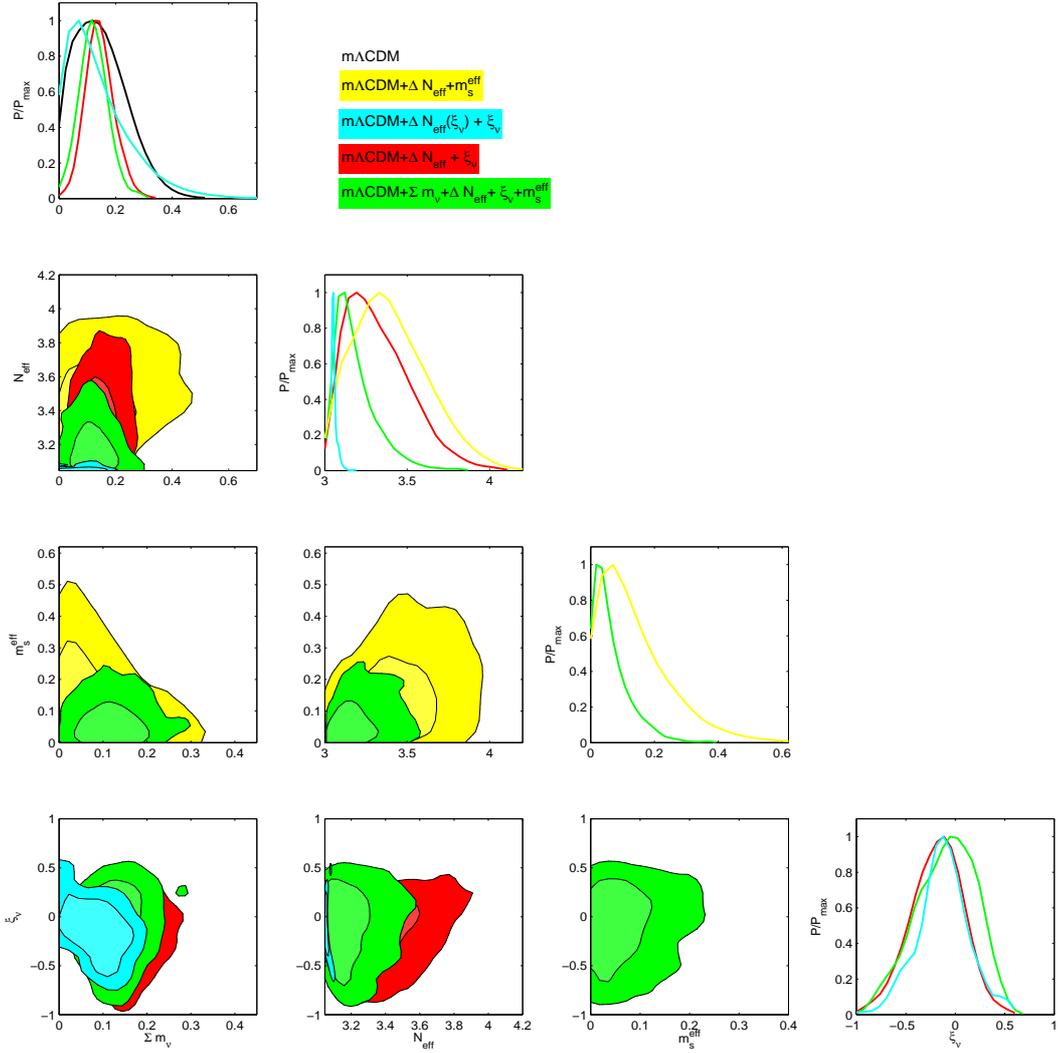}
\end{center}
\caption{The marginalized posterior distributions for $\Sigma m_{\nu}$, $N_{eff}$, $m^{eff}_s$ and
$\xi_{\nu}$
obtained from the fits of different extentions of the base $\Lambda$CDM model
to the {\sc Planck}+WP+highL+BAO+lensing dataset.}
\label{mactiv}
\end{figure}
\begin{figure}
\begin{center}
\includegraphics[height=10cm,width=16cm]{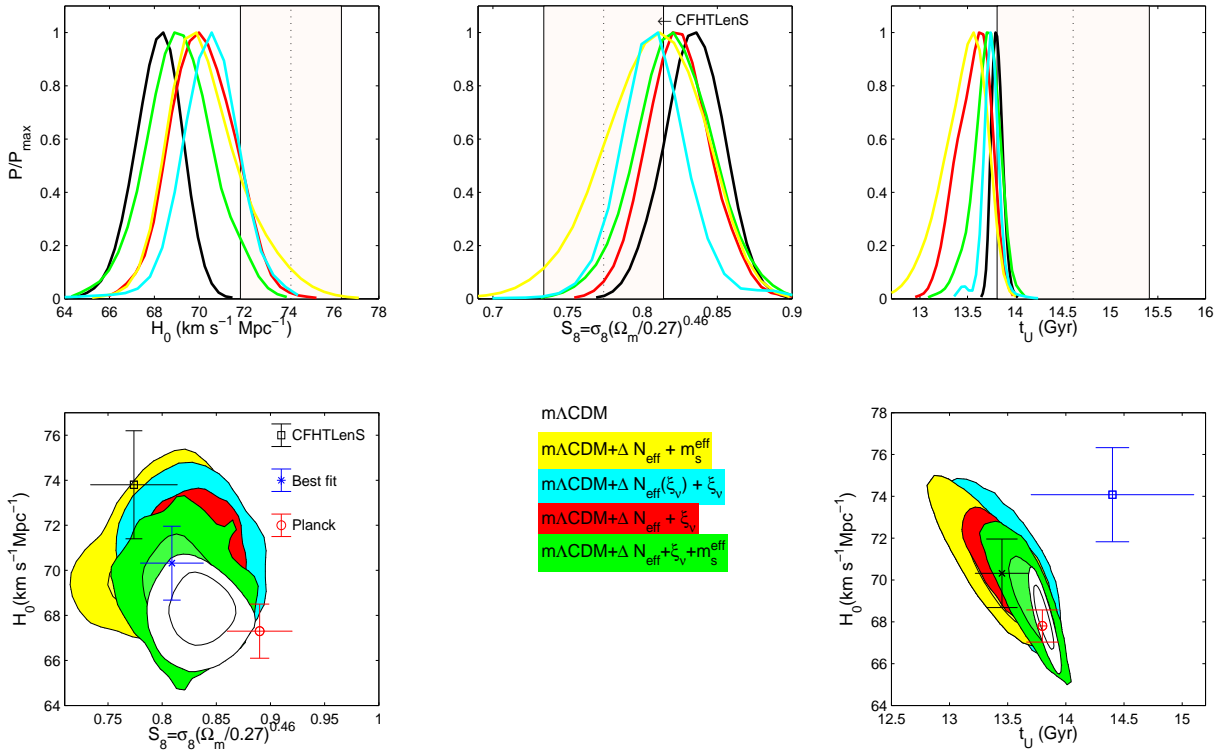}
\end{center}
\caption{Top: The marginalized posterior probability distributions from the fits of 
different extentions of the base $\Lambda$CDM model
to the {\sc Planck}+WP+highL+BAO+lensing dataset vs.   
local measurements of $H_0$ and $t_U$ and the determination of $S_8$ from CFHTLenS survey data (all indicated by bands).
Bottom: $S_8$ - $H_0$ and $t_U$ - $H_0$ joint confidence regions (at 68\% and 95\% CL)
from the same analysis.}
\label{tens_deg}
\end{figure}
\begin{figure}
\begin{center}
\includegraphics[height=10cm,width=10cm]{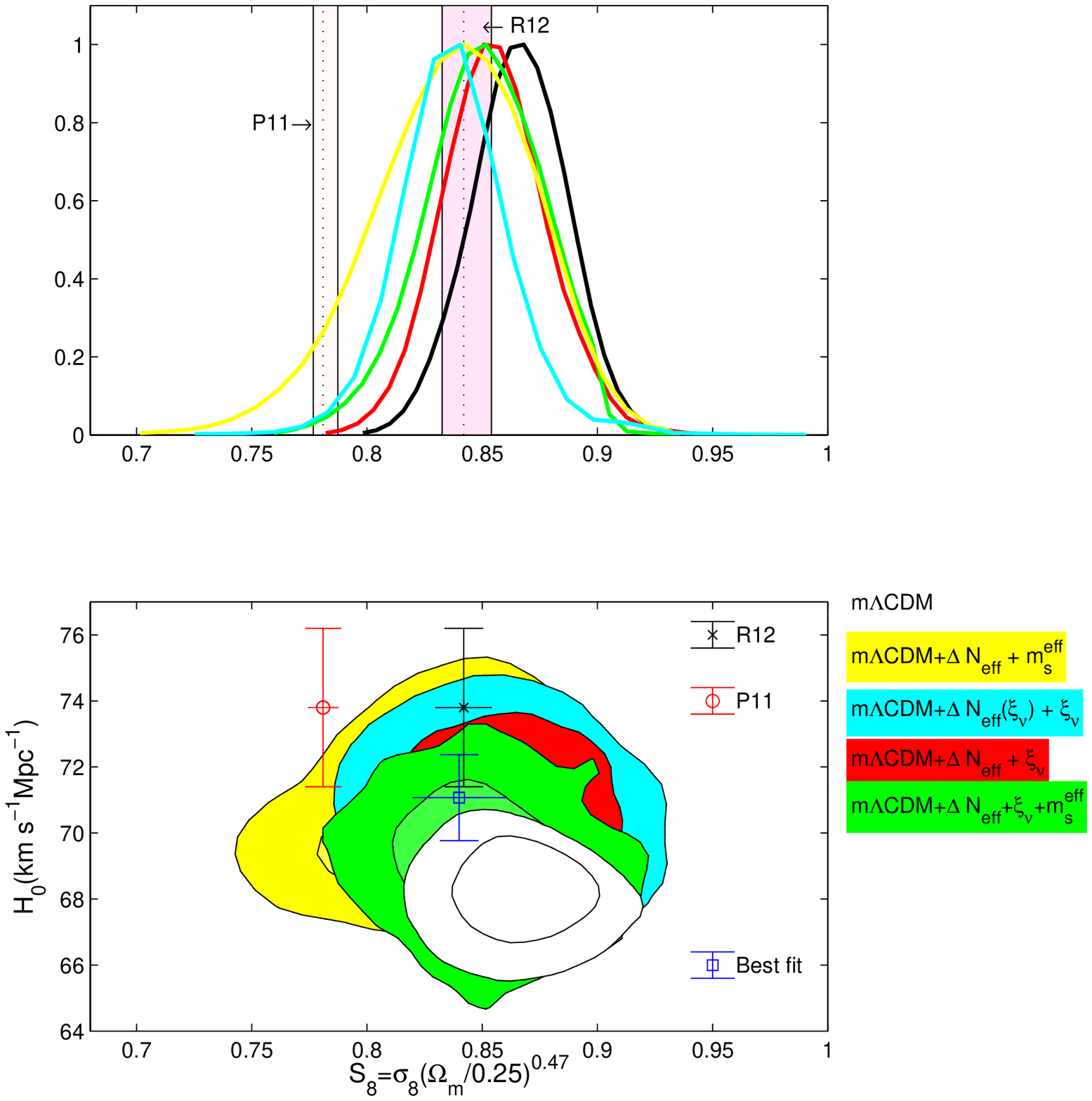}
\end{center}
\caption{Top: The marginalized posterior probability distributions for
$S_8$ obtained from the fits of different extentions of the base $\Lambda$CDM model
to the {\sc Planck}+WP+highL+BAO+lensing dataset  vs. $S_8$ values obtained by using the cluster 
mass-observation relation  for two mass calibration offset values P11 \cite{Ade2011} and R12 
\cite{Rozo2012b} (indicated by bands).
Bottom: $S_8$ - $H_0$ joint confidence regions (at 68\% and 95\% CL)
obtained from the same analysis.}
\label{tens_deg_clust}
\end{figure}
In this section we explore the extensions of the base $\Lambda$CDM model by considering  the leptonic
asymmetry and extra radiation energy density for the case of three degenerate massive neutrino species. 
Our main results are summarized in Table \ref{tab:neutrino_deg}. In
Figure \ref{mactiv} we present the marginalized posterior distributions for $\Sigma m_{\nu}$, $N_{eff}$, $m^{eff}_s$ and $\xi_{\nu}$ obtained from the fits of different extentions of the base $\Lambda$CDM model
to the {\sc Planck}+WP+highL+BAO+lensing dataset. In Figure \ref{tens_deg} we compare the marginalized posterior probability distributions 
and the  joint confidence regions obtained from our fits with   
the local measures of $H_0$ and $t_U$ and the determination of $S_8$ from CFHTLenS survey data.
A similar comparison is made in Figure \ref{tens_deg_clust} for $S_8$ values obtained by using the cluster 
mass-observation relation  for mass calibration offset values P11 and R12. \\
We start with $m\Lambda$CDM, the minimal extention of $\Lambda$CDM model.
Adding massive neutrinos to $\Lambda$CDM maintains the tension with
local measure of $H_0$ at 2.4$\sigma$  
but reduces the discrepancy with the determination of $S_8$ from 
CFHTLenS survey data and local measure of $t_U$ at about 1.3$\sigma$ and 0.84$\sigma$ respectively. 
The tension between $S_8$ values is even more reduced (at about $0.35\sigma$) 
when the cluster mass-observation relation is employed with the calibration offset value R12.
Allowing part of the matter to be composed by neutrinos with eV mass suppresses the growth of structures
below neutrino free-streaming scale, leading to a smaller value for $\sigma_8$. \\
The extension of $m\Lambda$CDM by allowing the presence of the extra relativistic
species and one thermal sterile neutrino alters the physical scales associated to CMB and BAO, broadening
the allowed ranges for $H_0$ and $S_8$ posterior distributions
to include  a larger overlap with the corresponding values from other measurements, 
as shown in Figure \ref{tens_deg} and Figure \ref{tens_deg_clust}.
As neutrinos with eV mass decouple when they are still relativistic ($T_{dec} \sim$ 2 MeV),
the main effect of including $\Delta N^{oth}_{eff} \ne 0$ is the change of relativistic energy density. This
changes the redshift of matter-radiation equality, $z_{eq}$, that affects the determination of
$\Omega_m h^2$ from CMB measurements because of its linear dependence on $N_{eff}$ \cite{Popa2008}:
\begin{eqnarray}
1+z_{eq}=\frac{\Omega_m h^2}{\Omega_{\gamma}h^2} \frac{1}{1+0.2271N_{eff}}\,,
\end{eqnarray}
where $\Omega_{\gamma}h^2$=2.469$\times 10^{-5}$ is the photons energy density for
$T_{cmb}$ = 2.725~K.  As consequence, $N_{eff}$ and $\Omega_m h^2$ are correlated,
with the width of degeneracy line given by the uncertainty in the determination of $z_{eq}$.
Moreover, sterile neutrino contributes to  the increase of matter energy density
with $\Omega_s h^2 = m^{eff}_{s}$/(94.1 eV), suppressing the grow of structures below
its free streaming scale.
We find for this model that the discrepancies with the local measure of $H_0$ and the determination of $S_8$ from 
CFHTLenS survey data are alleviated at 1.35$\sigma$ and 1.11$\sigma$ respectively. 
We find practicaly no tension between $S_8$ values ($< 0.14\sigma)$ when the cluster mass-observation relation is employed with the calibration offset value R12.
However, the $N_{eff}$ value is disfavored at 1.6$\sigma$ by the SM value $N_{eff}$=3.046. \\
The non-zero neutrino chemical potential augments the extra energy density due to unknown processes by $\Delta N_{eff}(\xi_{\nu})$ as given in Equation (\ref{delta_neff_zeta}).
This imply a larger expansion rate of the Universe, an earlier weak process freeze out
with a higher value for the neutron to proton density ratio, and thus a larger value of
$Y_P$ . On the other hand, a non-zero value of the electron neutrino chemical potential, $\xi_{\nu_e}$, shifts the neutron-proton beta equilibrium, leading to a larger variation of $Y_P$. 
As the leptonic asymmetry leads to changes in neutrino free-streaming length and neutrino Jeans mass due to the increase of the neutrino velocity dispersion, the values of $\Sigma m_{\nu}$ and $m^{eff}_s$ are
decreased when compared to $\xi_{\nu}=0$ case. This increases the tension between cosmological and short baseline (SBL) neutrino oscillations data \cite{Giunti} that favors a sterile neutrino with the mass of $\sim$ 1 eV. \\
\begin{figure}
\begin{center}
\includegraphics[height=6cm,width=16cm]{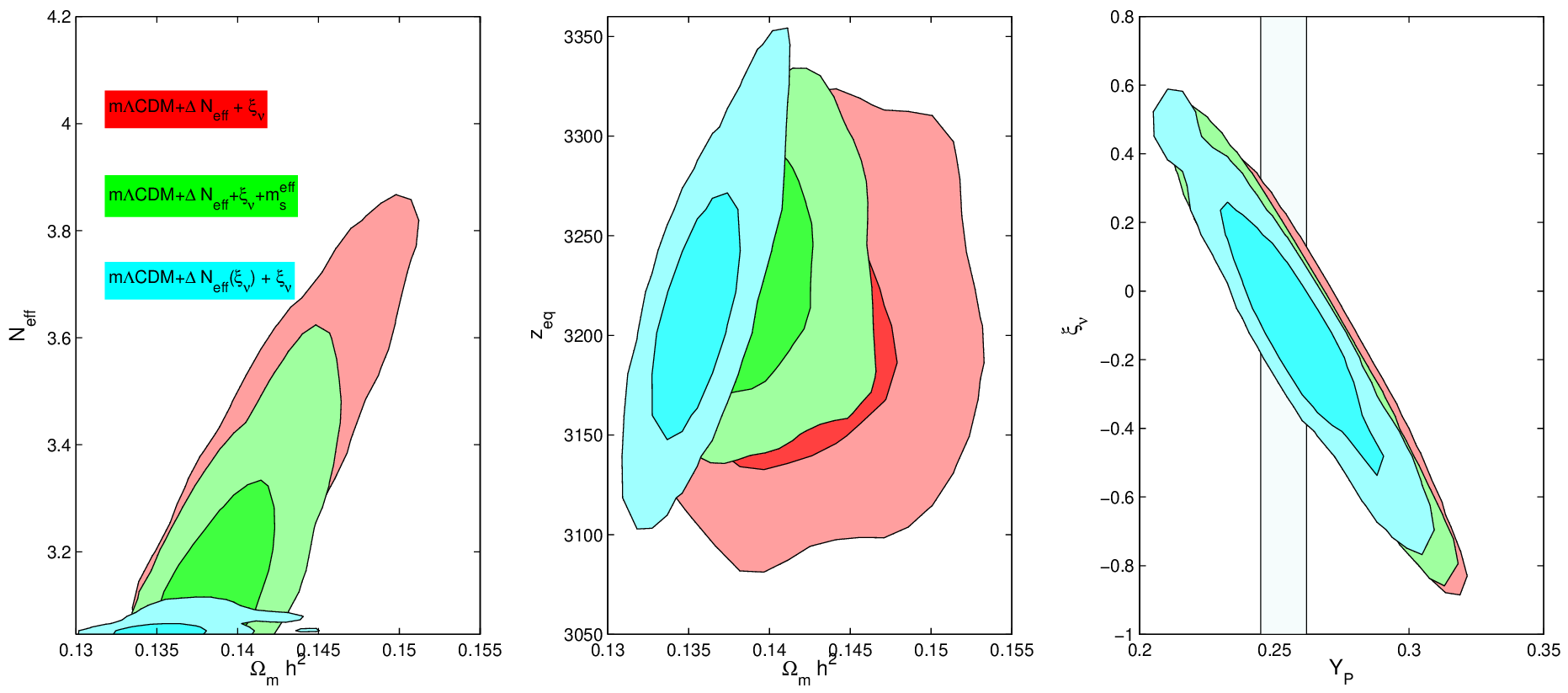}
\end{center}
\caption{The joint two-dimensional marginalized probability distributions (68$\%$ and
95$\%$ CL) in $\Omega_m h^2$ - $N_{eff}$, $\Omega_m h^2$ - $z_{eq}$, and $Y_P$ -$\xi_{\nu}$ planes
for the extensions of $m\Lambda$CDM models involving leptonic asymmetry. The vertical band shows the BBN
observational bounds (68$\%$ CL) on $^4$He primordial abundance \cite{Aver}.}
\label{zeq_Neff}
\end{figure}
In particular, we find for the extension of the $m\Lambda$CDM with $N^{oth}_{eff}=0$ and $m^{eff}_s=0$ priors
that the best fit values for $H_0$ and $t_U$ are in agreement with their local determinations at 1.3$\sigma$ and 1.12$\sigma$ respectively, while $S_8$ is in agreement with its determination from CFHTLenS survey data at 0.93$\sigma$ and 
with the prediction of cluster mass-observation relation with the calibration offset value R12 at 0.35$\sigma$.
The value of $N_{eff}$ is in agreement with the SM value $N_{eff}$=3.046 within 0.66$\sigma$. Also, for this model we find the best fit value of primordial $^4$He mass fraction
in agreement within 0.27$\sigma$ with the BBN observational constraint \cite{Aver}. \\
Figure \ref{zeq_Neff} presents the joint two-dimensional marginalized probability distributions in $\Omega_m h^2-N_{eff}$,
$\Omega_m h^2-z_{eq}$, and $Y_P-\xi_{\nu}$ planes
for the extensions of $m\Lambda$CDM models involving leptonic asymmetry.
When we transform $N_{eff}$ axis of the left panel to $z_{eq}$ axis from the middle panel we observe a
strong correlation between $z_{eq}$ and $\Omega_mh^2$ only for the model with $N^{oth}_{eff}=0$ and $m^{eff}_s=0$ priors.
As the anisotropic stress of neutrinos leaves distinct signatures in the CMB power
spectrum which are correlated with $\Omega_m h^2$ , we conclude that in this case we
observe the effect of neutrino anisotropic stress rather than the effect of $z_{eq}$ change, as in the case of the other two models. 

We conclude that the current cosmological data favor
the leptonic asymmetric extension of $m\Lambda$CDM cosmological model over one with
additional sterile neutrino species.
\begin{table}[tb]
\caption{The table shows the mean values and the absolute errors on the
main cosmological parameters obtained from the 	fits of different extentions
of the $\Lambda$CDM model discussed in the text with {\sc Planck}+WP+highL+BAO+lensing
dataset for degenerate massive neutrinos. For all parameters, except $\Sigma m_{\nu}$ and $m^{eff}_s$,
we quote the errors at $68\%$ CL.
For $\Sigma m_{\nu}$ and $m^{eff}_s$ we give the values of the $95\%$ upper limits.}
\begin{small}
\begin{center}
\begin{tabular}{|l|l|l|l|l|l|}
\hline
Model & $m\Lambda$CDM & $m\Lambda$CDM +& $m\Lambda$CDM +& $m\Lambda$CDM +&$m\Lambda$CDM +  \\
      &               &$\Delta N^{oth}_{eff}+m^{eff}_s$& $\Delta N_{eff}(\xi_{\nu})+\xi_{\nu}$&
      $\Delta N_{eff}+\xi_{\nu}$& $\Delta N_{eff}+\xi_{\nu}+m^{eff}_s$ \\
\hline
$\Sigma m_{\nu}$(eV)& $<$ 0.347 & $<$ 0.276& $<$ 0.168& $<$ 0.250 & $<$ 0.231\\
$ N_{eff}$& - & 3.402 $\pm$ 0.224& 3.058 $\pm$ 0.018 & 3.328 $\pm$ 0.193 & 3.189 $\pm$ 0.130\\
$\xi_{\nu}$& - & -&-0.125 $\pm$ 0.255&-0.186 $\pm$ 0.261 &-0.097 $\pm$ 0.298\\
$m^{eff}_s$(eV)& - & $<$ 0.424 & - & -  & $<$ 0.214  \\
$Y_P$& 0.24 &  0.253 $\pm$ 0.003& 0.259 $\pm$ 0.019& 0.268 $\pm$ 0.020& 0.259 $\pm$ 0.023\\
\hline
$H_{0}$(km s$^{-1}$Mpc$^{-1}$)&  68.11 $\pm$ 1.02& 70.32 $\pm$ 1.64 & 70.45 $\pm$ 1.33& 70.15 $\pm$ 1.36 & 68.96 $\pm$ 1.41\\
$t_{U}$(Gyr)& 13.81 $\pm$ 0.06 & 13.47 $\pm$ 0.18 & 13.74 $\pm$ 0.09&13.55 $\pm$ 0.17 &
13.71 $\pm$ 0.14 \\
$S_8$& 0.831 $\pm$ 0.021 & 0.824 $\pm$ 0.019 & 0.831 $\pm$ 0.027 &
0.824 $\pm$ 0.021 & 0.823 $\pm$ 0.025\\
\hline
\end{tabular}
\end{center}
\end{small}
\label{tab:neutrino_deg}
\end{table}
\subsubsection{Non-degenerate massive neutrinos}
\begin{figure}
\label{tens_deg_H}
\begin{center}
\includegraphics[height=10cm,width=16cm]{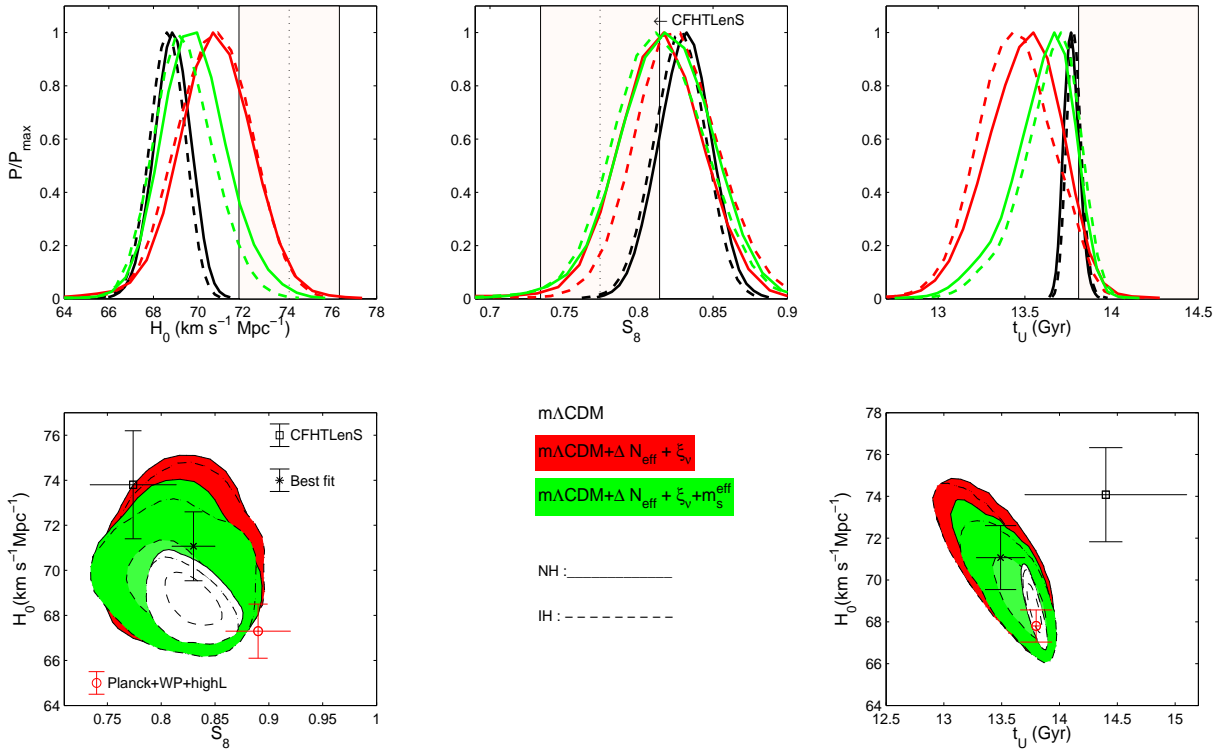}
\end{center}
\caption{The marginalized posterior probability distributions from the fits of 
different extentions of the base $\Lambda$CDM model 
to the {\sc Planck}+WP+highL+BAO+lensing dataset for normal (continuous lines) and inverted
(dashed lines) neutrino mass hierarchies vs.   
the local measurements of $H_0$ and $t_U$ and the determination of $S_8$ from CFHTLenS survey data (all indicated by bands).
Bottom: $S_8$ - $H_0$ and $t_U$ - $H_0$ joint confidence regions (at 68\% and 95\% CL)
from the same analysis.}
\label{tens_deg_H}
\end{figure}
\begin{figure}
\begin{center}
\includegraphics[height=10cm,width=10cm]{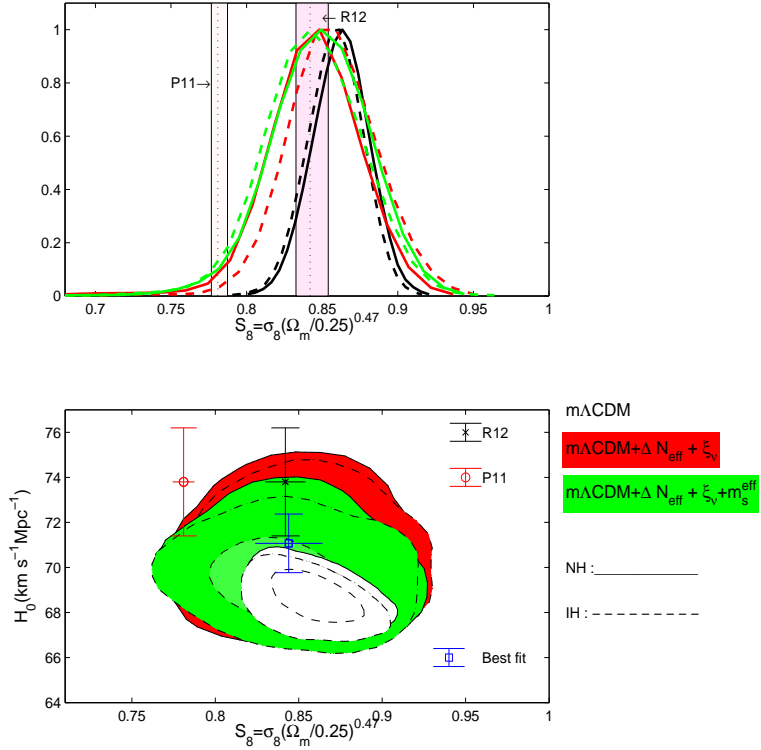}
\end{center}
\caption{Top: The marginalized posterior probability distributions for
$S_8$ obtained from the fits of different extentions of the base $\Lambda$CDM model
to the {\sc Planck}+WP+highL+BAO+lensing dataset for normal (continuous lines) and inverted
(dashed lines) neutrino mass hierarchies  vs. $S_8$ values obtained by using the cluster 
mass-observation relation  for two mass calibration offset values P11 \cite{Ade2011} and R12 
\cite{Rozo2012b} (indicated by bands).
Bottom: $S_8$ - $H_0$ joint confidence regions (at 68\% and 95\% CL)
obtained from the same analysis.}
\label{tens_deg_clust_H}
\end{figure}
\begin{figure}
\begin{center}
\includegraphics[height=7cm,width=16cm]{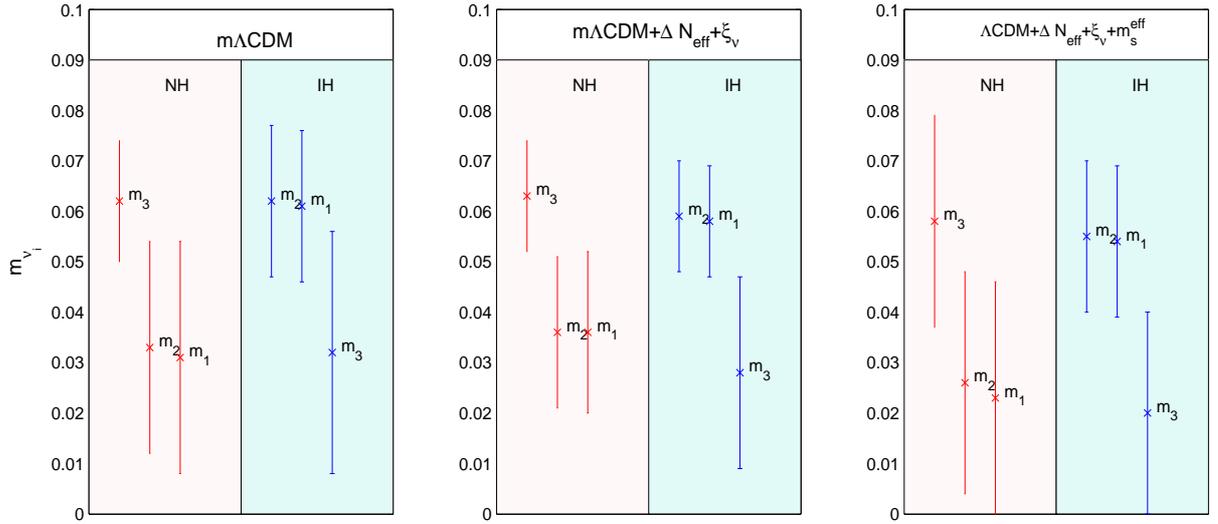}
\end{center}
\caption{Neutrino mass eigenstates ordered in the normal hierarchy (NH: $m_1 < m_2 < m_3$) and
in the inverted one (IH: $m_3 < m_1 < m_2$)
obtained from the fit of  different extentions of the base $\Lambda$CDM model
to the {\sc Planck}+WP+highL+BAO+lensing dataset (1$\sigma$ error bars).}
\label{H}
\end{figure}
\begin{table}[tb]
\caption{The table shows the mean values and the absolute errors (68$\%$ CL) on the
main cosmological parameters obtained from the fits of different extentions
of the $\Lambda$CDM model discussed in the text with {\sc Planck}+WP+highL+BAO+lensing dataset,
for neutrino normal mass hierarchy (NH) and the inverted one (IH).}
\begin{small}

\vspace{0.2cm}
\begin{tabular}{|l|lll|lll|}
\hline
          &     &${\bf NH}$&   &     & ${\bf IH}$ & \\
Model     & $m\Lambda$CDM  &$m\Lambda$CDM+ & $m\Lambda$CDM+& $m\Lambda$CDM&$m\Lambda$CDM+&$m\Lambda$CDM+  \\
          & & $\Delta N_{eff}+\xi_{\nu}$& $\Delta N_{eff}+\xi_{\nu}$
          & & $\Delta N_{eff}+\xi_{\nu}$& $\Delta N_{eff}+\xi_{\nu}$ \\
          & &   & + $m^{eff}_s$& & & +$m^{eff}_s$\\
\hline
$\Sigma m_{\nu}$(eV)& 0.125 $\pm$ 0.057 & 0.132 $\pm$ 0.048 & 0.100 $\pm$ 0.032 & 0.154 $\pm$ 0.054 & 0.145 $\pm$  0.042 &  0.128 $\pm$ 0.031 \\
$m_1$(eV) & 0.031 $\pm$ 0.023 & 0.034 $\pm$ 0.018 & 0.023 $\pm$ 0.023
& 0.061 $\pm$ 0.015 & 0.058 $\pm$ 0.011 & 0.054 $\pm$ 0.015\\
$m_2$(eV) & 0.033 $\pm$ 0.021 & 0.036 $\pm$ 0.017 & 0.026 $\pm$ 0.022
& 0.062 $\pm$ 0.015 & 0.059 $\pm$ 0.011 & 0.055 $\pm$ 0.015\\
$m_3$(eV) & 0.062 $\pm$ 0.012 & 0.063 $\pm$ 0.011 & 0.058 $\pm$ 0.011
& 0.032 $\pm$ 0.024 &0.028 $\pm$ 0.019 & 0.020 $\pm$ 0.021  \\
$ N_{eff}$&- & 3.47 $\pm$ 0.28 & 3.21 $\pm$0.14 & - & 3.46 $\pm$ 0.28& 3.26 $\pm$ 0.21\\
$\xi_{\nu}$& - & -0.073 $\pm$ 0.311 & -0.035 $\pm$ 0.290 & - & 0.121 $\pm$ 0.362 &
0.024 $\pm$ 0.34\\
$m^{eff}_s$(eV)&  - & - & 0.076 $\pm$ 0.070 & - & - & 0.073 $\pm$ 0.0.071\\
$Y_P$&    0.24 & 0.251 $\pm$ 0.019 & 0.255 $\pm$ 0.026 & 0.24 & 0.246 $\pm$ 0.026 &
0.254 $\pm$ 0.026\\
\hline
$H_{0}$& 68.83 $\pm$ 0.80 & 71.07 $\pm$ 1.53 & 70.01 $\pm$ 0.98 & 68.59 $\pm$ 0.76 & 70.14 $\pm$ 1.55 & 69.09 $\pm$ 1.14\\
$t_U$(Gyr)& 13.77 $\pm$ 0.04 & 13.49 $\pm$ 0.21 & 13.67 $\pm$ 0.12 & 13.78 $\pm$ 0.04 &
13.49 $\pm$ 0.22 & 13.66 $\pm$ 0.16 \\
$S_8$&0.829 $\pm$ 0.020& 0.830 $\pm$ 0.022 & 0.828 $\pm$ 0.026 & 0.827 $\pm$ 0.016 & 0.842 $\pm$ 0.022& 0.823 $\pm$ 0.027\\
\hline
\end{tabular}
\end{small}
\label{tab:neutrino_H}
\end{table}
In this section we explore the extensions of the base $\Lambda$CDM model
by considering the neutrino mass ordering.
For the models discussed in the previous section we consider
three species of non-degenerate massive neutrinos with the mass eigenvalues
$m_1$, $m_2$ and $m_3$ ordered in normal hierarchy (NH: $m_1 < m_2 < m_3$) and
in inverted one (IH: $m_3 < m_1 < m_2$).
The neutrino total mass for each hierarchy is given by:
\begin{eqnarray}
{\rm NH :}\,\, \Sigma m_{\nu} & = &  m_{min} +
                  \sqrt{\Delta m^2_{21}+ m^2_{min}} +
                  \sqrt{\Delta m^2_{31}+ m^2_{min}}\,,
\,\,\, m_{min}=m_1  \,, \\
{\rm IH : } \,\, \Sigma m_{\nu} & =&  \sqrt{m^2_{min}- \Delta m^2_{32}-\Delta m^2_{21}} + \sqrt{m^2_{min} - \Delta m^2_{32}} +m_{min} \,,
\,\,\, m_{min}=m_3 \,,\nonumber
\end{eqnarray}
where $\Delta m_{21}$, $\Delta m_{31}$, $\Delta m_{32}$
are the neutrino mass squared differences.
We take the central values of neutrino mass squared differences obtained from the global fit of neutrino mixing parameters 
 \cite{Maltoni}:
\begin{eqnarray}
\Delta m^2_{21}=7.5 \times 10^{-5}{\rm eV^2} \,,\,
\Delta m^2_{31}=2.47\times 10^{-3}{\rm eV^2} \,,\,
\Delta m^2_{32}=-2.43 \times 10^{-3}{\rm eV^2}. \nonumber
\end{eqnarray}
Table~\ref{tab:neutrino_H} presents the mean values and the absolute errors (68$\%$ CL) on the
main cosmological parameters obtained from the fits of different extentions
of the $\Lambda$CDM model discussed in the text with {\sc Planck}+WP+highL+BAO dataset,
for neutrino normal and inverted mass hierarchies.
In Figure \ref{tens_deg_H} we compare the marginalized posterior probability distributions 
and the  joint confidence regions obtained from our fits with   
the local measures of $H_0$ and $t_U$ and the determination of $S_8$ from CFHTLenS survey data 
for neutrino normal mass hierarchy and the inverted one.
A similar comparison is made in Figure \ref{tens_deg_clust_H} for $S_8$ values obtained by using the cluster 
mass-observation relation for mass calibration offset values P11 and R12.

The results presented clearly demonstrate the preference of cosmological data 
for the case of cosmological model involving neutrino chemical potential ($\xi_{\nu} \ne 0$) and three massive neutrino species ($m^{eff}_s$=0) with direct mass hierarchy,
that favors a total neutrino mass $\Sigma m_{\nu} \sim 0.132$ eV ($\sim 2 \sigma$ statistical evidence). For this model we find that the best fit values for 
$H_0$ and $t_U$ are in agreement with their local measures within 1.1$\sigma$ and 1.3$\sigma$ respectively,
while $S_8$ is in agreement with its determination from CFHTLenS survey data at 1.2$\sigma$ and 
with the prediction of cluster mass-observation relation with the calibration offset value R12 at 0.05$\sigma$.
The value of $N_{eff}$ agrees with its SM value within 1.3$\sigma$. \\ 
Figure \ref{H} presents the neutrino mass eigenstates ordered in normal hierarchy (NH: $m_1 < m_2 < m_3$) and
in the inverted one (IH: $m_3 < m_1 < m_2$) as obtained from the fit of  different extentions of the base $\Lambda$CDM model to the {\sc Planck}+WP+highL+BAO+lensing dataset.

We conclude that the current cosmological data favor
the leptonic asymmetric extension of $m\Lambda$CDM cosmological model and normal neutrino mass hierarchy over the models  with additional sterile neutrino species and/or inverted neutrino mass hierarchy.

\section{Conclusions}

Recently, the {\sc Planck} satellite found a larger and most precise value of
the matter energy density, that impacts on the present values of
other cosmological parameters such as the Hubble constant $H_0$, the present cluster abundances 
$S_8$ and the age of the Universe $t_U$.
The existing tension between {\sc Planck} determination of these parameters in the frame
of the base $\Lambda$CDM model and their determination from other measurements generated lively discussions,
one possible interpretation being that some sources of systematic errors in cosmological measurements are not completely understood \cite{Planck1}. 
An alternative interpretation is related to the fact that the CMB observations, 
that  probe the high redshift Universe are interpreted in terms of cosmological parameters at present time  by extrapolation within the base $\Lambda$CDM model that can be inadequate or incomplete. \\
In this paper we quantify this tension
by exploring several extensions of the base $\Lambda$CDM model that include the leptonic asymmetry
restricted in the form of neutrinos from the requirement of universal electric neutrality. \\
We set bounds on the radiation content of the Universe and neutrino properties by using the latest cosmological measurements ({\sc Planck}+WP+highL+BAO+lensing),
imposing also self-consistent BBN constraints on the primordial helium abundance, which proved to be important in
the estimation of cosmological parameters \cite{Planck1,Aver}. We consider lepton asymmetric cosmological models  parametrized by the neutrino total mass of three massive neutrino species $\Sigma m_{\nu}$,
neutrino degeneracy parameter $\xi_{\nu}$, the variation of the extra relativistic degrees of freedom $\Delta N_{eff}$, and one thermally distributed sterile neutrino with the mass $m_s$.

From the analysis of fundamental MCMC parameters we obtain
the posterior probability distributions for $H_0$, $S_8$ and $t_U$
that we then compare to the corresponding values from other measurements.
We study the consistency and cosmological implications of the leptonic asymmetry considering: i) degenerate massive neutrinos and ii) non-degenerate massive neutrinos with the masses ordered in normal hierarchy (NI) and
inverted one (IH) and neutrino mass squared differences obtained from the global
fit of neutrino mixing parameters \cite{Maltoni}.\\
For all cosmological asymmetric models studied
we find the preference of cosmological data for smaller values of $\Sigma m_{\nu}$ and $m_s$ when compared with
$\xi_{\nu}=0$ case. This increases the tension between cosmological and short baseline neutrino oscillation data
that favors a sterile neutrino with the mass of $\sim$1 eV.\\
For the case of degenerate massive neutrinos, we find that the discrepancies with the local determinations of $H_0$, and $t_U$ are alleviated at $\sim 1.3\sigma$ level while $S_8$ is in agreement with its determination from CFHTLenS survey data at $\sim 1 \sigma$ and with the prediction of cluster mass-observation relation with calibration offset value R12 at $\sim 0.5\sigma$. This conclusion is valid for all extensions of the base $\Lambda$CDM  with $\xi_{\nu} \ne 0$, except the model with $m_s \ne 0$ prior for which the tension is still present (at $\sim 2.5 \sigma$ level).\\
We also find the preference of cosmological data 
for the cosmological model involving neutrino chemical potential ($\xi_{\nu} \ne 0$) and three massive neutrino species ($m^{eff}_s$=0) with direct mass hierarchy,
that favors a total neutrino mass $\Sigma m_{\nu} \sim 0.132$ eV ($\sim 2 \sigma$ statistical evidence). For this model we find that the best fit values for 
$H_0$ and $t_U$ are in agreement with their local measures within 1.1$\sigma$ and 1.3$\sigma$ respectively,
while $S_8$ is in agreement with its determination from CFHTLenS survey data at 1.2$\sigma$ and 
with the prediction of cluster mass-observation relation with the calibration offset value R12 at 0.05$\sigma$.
The value of $N_{eff}$ agrees with its SM value within 1.3$\sigma$. \\ 

We conclude that the current cosmological data favor
the leptonic asymmetric extension of the base $\Lambda$CDM model and normal neutrino mass hierarchy over the models  with additional sterile neutrino species and/or inverted neutrino mass hierarchy.

\vspace{1cm}
{\bf Acknowledgments}\\ \\
This work  was supported by CNCSIS Contract 82/2013 and by ESA/PECS Contract C98051.\\
We also acknowledge the use of the GRID computing system facility at the Institute of
Space Science Bucharest and would like to thank the staff working there.

 \end{document}